\documentclass[
aip,
reprint,
superscriptaddress,
longbibliography,
showkeys,
amsmath,amssymb,
aps,
pre, 
]{revtex4-1}

\usepackage{graphicx}
\usepackage[caption=false]{subfig} 

\usepackage{dcolumn}
\usepackage{bm}

\newcommand{\ee}{\end{equation}} 
\newcommand{\be}{\begin{equation}}

\usepackage[mathscr]{euscript}  
\usepackage{xcolor}  

\usepackage{tcolorbox}
\usepackage{comment}
\usepackage{float}
\usepackage{soul}

\usepackage{array}

\makeatletter
\newsavebox{\@brx}
\newcommand{\llangle}[1][]{\savebox{\@brx}{\(\m@th{#1\langle}\)}%
  \mathopen{\copy\@brx\kern-0.5\wd\@brx\usebox{\@brx}}}
\newcommand{\rrangle}[1][]{\savebox{\@brx}{\(\m@th{#1\rangle}\)}%
  \mathclose{\copy\@brx\kern-0.5\wd\@brx\usebox{\@brx}}}
\makeatother

\begin{document}

\preprint{APS/123-QED}

\title{Harnessing non-equilibrium forces to optimize work extraction}

\author{Kristian St\o{}levik Olsen}
\thanks{\textit{Correspondence: kristian.olsen@hhu.de}}
\affiliation{Institut für Theoretische Physik II - Weiche Materie, Heinrich-Heine-Universität Düsseldorf,\\ D-40225 Düsseldorf, Germany}

\author{R\'{e}mi Goerlich}
\affiliation{Raymond \& Beverly Sackler School of Chemistry, Tel Aviv University,\\ Tel Aviv 6997801, Israel}

\author{Yael Roichman}
\affiliation{Raymond \& Beverly Sackler School of Chemistry, Tel Aviv University,\\ Tel Aviv 6997801, Israel}
\affiliation{Raymond \& Beverly Sackler School of Physics \& Astronomy,Tel Aviv University,\\ Tel Aviv 6997801, Israel}

\author{Hartmut L\"{o}wen}
\affiliation{Institut für Theoretische Physik II - Weiche Materie, Heinrich-Heine-Universität Düsseldorf,\\ D-40225 Düsseldorf, Germany}

\begin{abstract}
{\color{black} 

While optimal control theory offers effective strategies for minimizing energetic costs in noisy microscopic systems over finite durations, a significant opportunity lies in exploiting the temporal structure of non-equilibrium forces. We demonstrate this by presenting exact analytical forms for the optimal protocol and the corresponding work for any driving force and protocol duration. We also derive a general quasistatic bound on the work, relying only on the coarse-grained, time-integrated characteristics of the applied forces. Notably, we show that the optimal protocols often automatically act as information engines that harness information about non-equilibrium forces and an initial state measurement to extract work. These findings chart new directions for designing adaptive, energy-efficient strategies in noisy, time-dependent environments, as illustrated through our examples of periodic driving forces and active matter systems. By exploiting the temporal structure of non-equilibrium forces, this largely unexplored approach holds promise for substantial performance gains in microscopic devices operating at the nano- and microscale.}
\end{abstract}


\maketitle


\section*{Introduction}\label{sec:intro}
{ \color{black}

Over two centuries ago, the development of thermodynamics laid the foundation for the Industrial Revolution. In recent decades, major advances—particularly through the development of stochastic thermodynamics—have extended thermodynamic principles to microscopic systems, where thermal fluctuations play a dominant role \cite{seifert2012stochastic,ciliberto2017experiments,van2013stochastic,sekimoto1998langevin}. This emerging framework enables us to rigorously address two central challenges: how to optimally control small-scale processes under constraints of accuracy, speed, and minimal energy expenditure; and how to efficiently harvest energy from strongly fluctuating, far-from-equilibrium environments.

Harvesting energy from nonequilibrium forces and fluctuations has already proven successful in a variety of macroscopic technologies, including wave-energy converters that utilize oscillatory forces \cite{falcao2010wave}, piezoelectric devices powered by biomechanical deformation \cite{azimi2021self,panda2022piezoelectric,lucente2025optimal}, and wearables that generate energy from human motion \cite{anwar2021piezoelectric}. At microscopic scales, this principle may be even more consequential, as both biological and synthetic systems routinely operate in dynamic, out-of-equilibrium conditions.

There is growing evidence that non-equilibrium fluctuations are not just unavoidable noise, but can be harnessed as a resource. For example, biological systems such as molecular motors perform micro- and nanoscale tasks with remarkable efficiency despite operating under noisy and driven conditions \cite{toyabe2011thermodynamic,ariga2021noise}. Likewise, non-equilibrium stochastic engines have in some cases been shown to outperform their equilibrium counterparts by exploiting fluctuations \cite{movilla2021energy,miangolarra2022geometry,ventura2024refined,saha2023information,abdoli2025enhanced,chor2023many}. A deeper understanding of these effects may prove crucial for the future design of efficient, robust small-scale engines \cite{martinez2017colloidal}.

As technological innovation continues to push the boundaries of miniaturization, identifying the fundamental limits of these processes—and designing control strategies that minimize energetic and temporal costs—has become essential for the optimal operation and design of next-generation microscopic machines.

Of particular interest is the development of optimal protocols—strategies for varying control parameters over time to drive a system between two states while minimizing costs such as energy, dissipation, or duration \cite{schmiedl2007optimal,gomez2008optimal,aurell2011optimal,bechhoefer2021control}. Recent advances have extended these concepts to more complex, non-homogeneous environments, including disordered media \cite{khadem2022stochastic,venturelli2024stochastic}, stochastic resetting processes \cite{gupta2020work}, and viscoelastic backgrounds \cite{loos2023universal}. Optimal protocols have also been studied in systems involving multiple or constrained control parameters \cite{plata2019optimal}. Additionally, approaches based on information geometry and thermodynamic metrics have yielded broad, unifying insights into optimal control in far-from-equilibrium systems \cite{van2023thermodynamic}.

\begin{figure*}[t!]
    \centering
    \includegraphics[width=15cm]{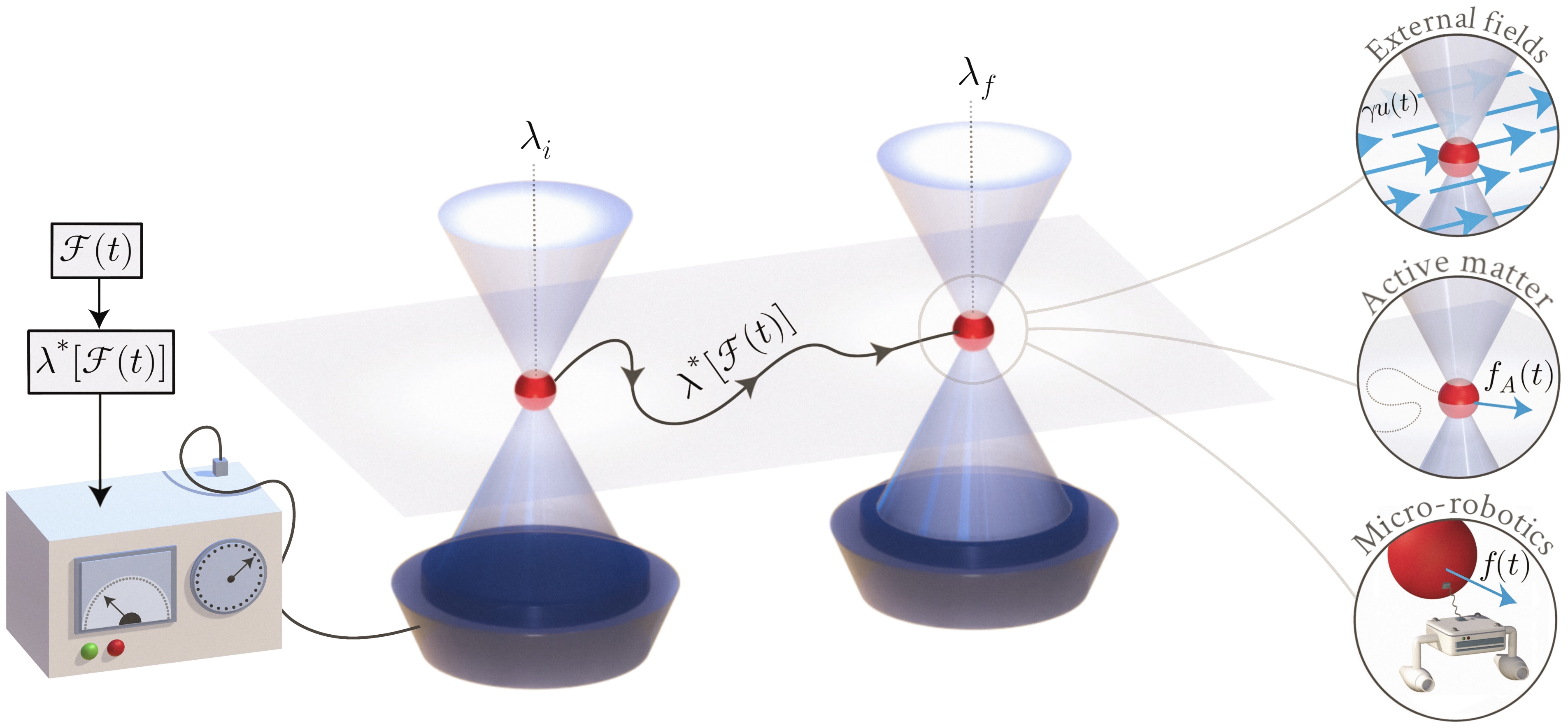}
    \caption{Optimal protocols for particle transport are derived for particles under the influence of arbitrary time-dependent forces $\bm{\mathscr{F}}(t)$. Given knowledge about these forces, a harmonic trap with a movable center $\bm{\lambda}(t)$ is tuned by the controller, at the cost of thermodynamic work. The forces may represent external drives such as applied fields or flows, or internally general forces such as in active matter. The optimal transport problem can also be interpreted in terms of micro-robotics and cargo transport, where $\lambda^*(t)$ represents the trajectory of a micro-robot transporting a cargo that is exposed to non-equilibrium conditions. }
    \label{fig:setup}
\end{figure*}

Despite these advances, a critical question remains largely unexplored: how to exploit the temporal structure of forces and fluctuations in non-equilibrium environments. Unlike equilibrium systems, which lack intrinsic time dependence, non-equilibrium settings often exhibit rich temporal features, such as characteristic timescales and frequency spectra, that may be harnessed for improved control and energy extraction. Indeed, generalized Landauer bounds suggest that excess work can be reduced by leveraging the information content separating a system’s non-equilibrium state from a reference equilibrium state \cite{esposito2011second,parrondo2015thermodynamics}.

In this work, we address this critical question by identifying optimal protocols that extract energy using information about time-dependent forces acting on a particle. Since forces induce translational displacements, it is natural to consider protocols that manipulate the trap via a translational degree of freedom, allowing the control strategy to adapt to force-induced effects. These forces may originate from external fields—such as fluid flows or electromagnetic fields—or arise internally, as in the case of self-propulsion forces driving active particles (see Fig.~\ref{fig:setup}). Taking a unified perspective, we present a general solution to the optimal control problem that applies broadly to such temporally driven systems. We illustrate this framework through the paradigmatic example of a particle confined in a harmonic potential, where the protocol governs the trap's position.

Considering a protocol $\bm{\lambda}(t)$ operating in a system subjected to time-dependent forces $\bm{\mathscr{F}}(t)$, we provide the exact form of the \emph{optimal} protocol for a desired operation: $\bm{\lambda}(t) = \bm{\lambda}[t, \bm{\mathscr{F}}(t)]$. These optimal protocols naturally decompose into a force-independent equilibrium contribution, $\bm{\lambda}_\text{eq}(t)$, and a force-dependent non-equilibrium contribution, $\bm{\lambda}_\text{neq}(t) = \bm{\lambda}_\text{neq}[t, \bm{\mathscr{F}}(t)]$. We then derive an exact expression for the thermodynamic work, valid for arbitrary driving forces $\bm{\mathscr{F}}(t)$ and protocol durations, and establish a quasistatic bound on the maximum extractable work. In the slow-driving limit, the total work separates into three distinct contributions: \textbf{i)} an information-geometric term quantifying how information from an initial non-equilibrium state can be converted into work, \textbf{ii)} the work required to slowly drag a particle in the presence of time-averaged forces, and \textbf{iii)} additional work that can be extracted by responding to fast dynamical modes in the driving. We illustrate the general results through several applications, including particles driven by periodic forces and a broad class of time-dependent forces relevant to the field of active matter.

}

\section*{Results}

\subsection{Optimal control of a colloidal particle in a harmonic trap far from thermal equilibrium} 

To illustrate our approach, we examine a well-established model system: a particle confined within a harmonic potential, where the control protocol governs the position of the trap center.An object, under the combined effect of an external driving force $\bm{f}(t)$, a harmonic potential with stiffness $k$ and trap center $\bm{\lambda}(t)$, and a thermal bath at temperature $T$,  obeys  the Langevin equation
\begin{equation}\label{eq:lang}
    \gamma \dot {\bm{x}}(t) = - k [\bm{x}(t) - \bm{\lambda}(t) ] + \bm{f}(t) +  \sqrt{2 k_B T \gamma}\bm{\xi}(t),
\end{equation}
where $\bm{\xi}(t)$ is Gaussian white noise with $\langle\bm{\xi}(t)\rangle=0$ and $\langle\bm{\xi}(t)\bm{\xi}(t')\rangle=\delta(t-t')$. This model could represent a colloidal particle manipulated by optical tweezers, or serve as a toy model for a motion protocol $\bm{\lambda}(t)$ for a micro-bot with spring-like coupling to a cargo at position $\bm{x}(t)$. The forces $\bm{f}(t)$, which for now are arbitrary and may be deterministic or stochastic, generically drive the system out of equilibrium. We denote by $\bm{q}(t)\equiv \langle \bm{x}(t) \rangle$ the mean particle trajectory, where the average is taken with respect to the Gaussian noise as well as any possible stochastic effects in the forces. We also let $\bm{\mathscr{F}}(t) \equiv \langle \bm{f}(t)\rangle$ denote the averaged force. The mean particle trajectory can be obtained by averaging the equation of motion, yielding
\begin{equation}\label{eq:qeq}
    \gamma \dot {\bm{q}}(t) = - k [ \bm{q}(t) -\bm{\lambda}(t) ] + \bm{\mathscr{F}}(t).
\end{equation}
We emphasize that if the forces are deterministic and known with precision, $\bm{\mathscr{F}}(t) = \bm{f}(t)$. Furthermore, any stochastic effect\st{s} may be either inherent to the forces, such as in the case of active self-propulsion forces, or represent effective forces resulting from errors in experimental measurements or inference. { \color{black}Recent efforts in force inference have employed a variety of approaches, including machine learning, Bayesian methods, and information-theoretic techniques\cite{frishman2020learning,wang2019machine,bryan2020inferring,chmiela2017machine,turkcan2012bayesian,masson2009inferring,sturm2024learning,majhi2025decodingactiveforcefluctuations}. This underscores the importance of \emph{information-limited optimization}, where protocols are derived based on the \emph{perceived} driving forces—those accessible through coarse-graining, averaging, or measurement uncertainty. For instance, if the true forces $\bm{f}(t; \kappa)$ depend on a parameter $\kappa$ known only with finite precision, typically distributed according to a Gaussian prior, it is natural to work with effective forces $\bm{\mathscr{F}}(t) = \langle \bm{f}(t; \kappa) \rangle_\kappa$ averaged over the distribution of measured values. These effective forces $\bm{\mathscr{F}}(t)$, appearing in the averaged dynamics, reflect the information available to the controller and represent the reproducible driving conditions across repeated experiments or simulations.}

Controlling the position of the harmonic trap $V[\bm{x},\bm{\lambda}(t)] = \frac{1}{2} k \left[ {\bm{x}} -{\bm{\lambda}}(t)\right]^2$ with a protocol ${\bm{\lambda}}(t)$ comes at a thermodynamic cost, which is given by the mean work
\begin{equation} \label{eq:wdef}
     \mathscr{W}[\{{\bm{\lambda}}(t)\}_0^{t_\text{f}}] =\int_0^{t_\text{f}} dt \left\langle \dot {\bm{\lambda}}(t)\cdot \frac{\partial V[\bm{x}(t),\bm{\lambda}(t)]}{\partial \bm{\lambda}(t)} \right\rangle.
\end{equation}

Here averages are taken over noise and stochastic force realizations as well as over measurement errors. Under the boundary conditions $\bm{\lambda}(t_0) = \bm{\lambda}_i$ and to $\bm{\lambda}(t_\text{f}) = \bm{\lambda}_f$, we seek the optimal protocol $\bm{\lambda}^*(t)$ that minimizes the work performed over a fixed time interval. In the absence of external forces ($\bm{\mathscr{F}}  =0$), the optimal protocol is known to be linear, with symmetric discontinuous jumps at the very beginning and end of the protocol \cite{schmiedl2007optimal}. Through Eq. (\ref{eq:qeq}) and Eq. (\ref{eq:wdef}), the work can be written as
\begin{align}
        \mathscr{W} &=  \int_0^{t_\text{f}} dt \: \dot {\bm{\lambda}}(t) \cdot \left[ \gamma\dot {\bm{q}}(t) - \bm{\mathscr{F}}(t)\right]\\
&= \text{Boundary terms} + \int_0^{t_f} d t \mathscr{L}(t, \bm{q},\dot {\bm{q}})
\end{align} 
where we defined a Lagrangian $\mathscr{L}(t, \bm{q},\dot {\bm{q}}) = \dot {\bm{q}}(t) \cdot [ \gamma \dot {\bm{q}}(t) -  \bm{\mathscr{F}}(t) ]$. The corresponding Euler-Lagrange equation $ \gamma \ddot {\bm{q}}(t) = \dot{\bm{\mathscr{F}}}(t)/2$ can be solved with the aforementioned boundary conditions, from which both the optimal protocol, mean particle position and work can be calculated exactly. See the appendix for technical details.

Under general $\bm{\mathscr{F}}(t)$, we derive an optimal protocol which can be decomposed into two contributions
\begin{align}\label{eq:split}
    \bm{\lambda}_*(t) &=  \bm{\lambda}_\text{eq}(t)   + \bm{\lambda}_\text{neq}(t).
\end{align}
The first term is an equilibrium contribution $(\bm{\mathscr{F}}=0)$, and the second a non-equilibrium contribution $(\bm{\mathscr{F}}\neq 0)$ that is determined solely by the perceived drive and protocol duration. Respectively, these take the form
\begin{align}
    \bm{\lambda}_\text{eq}(t) &= \bm{q}_i + \frac{1+\omega t}{2+\omega t_\text{f}}(\bm{\lambda}_f-\bm{q}_i),\label{eq:lambdaeq}\\
\bm{\lambda}_\text{neq}(t) &= \int_0^{t} dt' \frac{\bm{\mathscr{F}}(t')}{2\gamma}  -   \frac{1+ \omega t}{2+\omega t_\text{f}}   \int_0^{t_\text{f}} dt' \frac{\bm{\mathscr{F}}(t')}{2\gamma} -  \frac{\bm{\bm{\mathscr{F}}}(t)}{2\gamma\omega},\label{eq:lambdaneq}
\end{align}
where $\omega = k/\gamma$ is the inverse relaxation timescale of the harmonic trap. In the free diffusive limit, $\bm{\mathscr{F}} =0$ we recover the protocol first obtained in Ref.\cite{schmiedl2007optimal} in the case of an initial equilibrium state with $\bm{q}_i =0$. In this part of the protocol, $\bm{\lambda}_\text{eq}(t)$, consists of a straight line as a function of time but with discontinuous jumps at the beginning and end of the protocol. The non-equilibrium contribution to the protocol $\bm{\lambda}_\text{neq}(t)$, containing the force,  depends only on the duration of the protocol, but not on the initial and final location of the protocol. Hence, Eq.(\ref{eq:split}) can be interpreted as the equilibrium protocol with superimposed corrections that compensate for the driving forces $\bm{\mathscr{F}}(t)$. During the protocol, the mean particle path is given by 
\begin{align}
     \bm{q}(t) &= \bm{q}_i + \left[ \frac{\bm{\lambda}_f-\bm{q}_i}{2 + \omega t_\text{f} } - \frac{ 1}{ 2 + \omega t_\text{f}} \int_0^{t_\text{f}} dt' \frac{\bm{\mathscr{F}}(t')}{2\gamma}\right] \omega t \nonumber \\
     &+ \frac{1}{2 \gamma}\int_0^{t} dt' \bm{\mathscr{F}}(t').
     \label{eq:pos}
\end{align}
We note that there are two contributions to the path; first, a linear contribution that depends on both the details of the dynamics and protocol, and a second potentially non-linear time-dependence coming from the last term of Eq.~\ref{eq:pos}.

{ \color{black} We emphasize that, just like in Ref.\cite{schmiedl2007optimal}, the protocol only ensures that the potential is at the final location $\bm{\lambda}_f$ at time $t_\text{f}$, without any constraints on particle location. One could in principle also constrain the particle location at the end of the protocol, leading to increased control, but at the cost of less energy extraction. In this relation between control and cost, we consider protocols that are able to extract maximal work from the non-equilibrium forces. The particle position at the end of the protocol $\bm{q}(t_\text{f})$ will, in the quasistatic regime, be given as}
\begin{equation}\label{eq:qtf}
    \lim_{t_\text{f}\to \infty} \bm{q}(t_\text{f}) = \bm{\lambda}_f +\lim_{t_\text{f}\to \infty} \frac{\overline{\bm{\mathscr{F}}}(t_\text{f})}{k},
\end{equation}
{\color{black} where $\overline{\bm{\mathscr{F}}}(t_\text{f})$ is the time-averaged force (see Eq. (\ref{eq:timeavg}) in the following). Hence, the final particle position will, even in the quasistatic regime, deviate from the target location, in contrast to equilibrium systems \cite{schmiedl2007optimal}. }Surprisingly, this deviation is not determined by the value of the force at the later stages of the protocol, but by the full time-averaged force since the initial time $t=0$. The memory of the full dynamics is a consequence of the way in which the optimization intertwines the forces, the protocol, and the particle position.

The work associated with the optimal protocol can be shown to take the form
\begin{align}\label{eq:fullwork}
    \mathscr{W} =&   \frac{1}{2} k(\bm{\lambda}_f -  \bm{q}(t_\text{f}) )^2
    -\frac{1}{2}k(\bm{\lambda}_i -  \bm{q}_i )^2   \\
    & + \left(  \omega\frac{\bm{\lambda}_f-\bm{q}_i}{2 + \omega t_\text{f} } - \frac{ \omega}{ 2 + \omega t_\text{f}} \int_0^{t_\text{f}} dt \frac{\bm{\mathscr{F}}}{2 \gamma} \right)^2 \gamma  t_\text{f}  \nonumber \\
   &- \frac{1}{\gamma}\int_0^{t_\text{f}} dt   \left(\frac{\bm{\mathscr{F}}}{2}\right)^2 \nonumber 
\end{align}
which is an exact result valid for arbitrary protocol durations $t_\text{f}$ and for arbitrary forces $\bm{\mathscr{F}}(t)$.

\subsection*{Quasistatic bound on work extraction}

In many cases, knowing the qausistatic limit is informative, as it provides bounds on the work exchanged. Whether this bound is positive (costing work) or negative (extracting work) and bounded or infinite is of high practical relevance.  

Here the quasistatic limit of Eq.(\ref{eq:fullwork}), $\mathscr{W}_{qs} = \lim_{t_\text{f}\to \infty}\mathscr{W} $, is useful in several ways. Firstly, it offers intuition behind the terms contributing to the work and provides physical insights into the mechanisms by which work is extracted from knowledge of the forces.
Secondly, the quasistatic limit may be relevant to slow but finite-time experiments. Since, for an optically trapped particle, the potential does not change shape during the protocol, the equilibrium free energy difference is zero. Consequently, in the quasistatic limit, only non-equilibrium effects contribute, arising either from non-equilibrium initial conditions captured by $\bm{q}_i$ or from non-equilibrium driving forces $\bm{\mathscr{F}}(t)$.
 Taking the slow limit ($t_f\to \infty$) of Eq. (\ref{eq:fullwork}) we find

\begin{align}
    \mathscr{W}_\text{qs} =& \underbrace{- \frac{1}{2} k (\bm{\lambda}_i- \bm{q}_i)^2}_{\mathscr{W}_{i}}       \underbrace{- (\bm{q}_f-\bm{q}_i)  \lim_{t_\text{f}\to \infty} \overline{\bm{\mathscr{F}}} (t_\text{f})}_{\mathscr{W}_{ta}}\nonumber \\
    &\underbrace{- \frac{1}{4 \gamma }  \lim_{t_\text{f}\to \infty}  t_\text{f} \text{Var}(\bm{\mathscr{F}};t_\text{f})}_{\mathscr{W}_{d}}, \label{eq:Wqs}
\end{align} 
where we used the time-averaged mean and variance 
\begin{align} \label{eq:timeavg}
    \overline{\bm{\mathscr{F}}}(t_\text{f}) &=\frac{1}{t_\text{f}}\int_0^{t_\text{f}} dt'\bm{\mathscr{F}}(t'),  \\
    \text{Var}(\bm{\mathscr{F}};t_\text{f}) &= \frac{1}{t_\text{f}}\int_0^{t_\text{f}} dt'\bm{\mathscr{F}}^2(t')  -\left[\frac{1}{t_\text{f}}\int_0^{t_\text{f}} dt'\bm{\mathscr{F}}(t')\right]^2 .
\end{align}

Before we interpret each term in the quasistatic work, it is worth emphasizing the simplicity of this result. All terms in Eq.(\ref{eq:Wqs}) may be calculated directly through the forces acting on the free particle $\bm{\mathscr{F}}(t)$ in addition to the fixed boundary conditions of the protocol $\{\bm{\lambda}_i,\bm{\lambda}_f\}$. Hence, this formula can be applied to a wide range of systems without having to go through the optimization procedure explicitly. Furthermore, in the quasistatic limit, the work is determined solely by the first two time-integrated cumulants, rendering higher-order fluctuations irrelevant.

In Eq.~(\ref{eq:Wqs}), the first term is determined by the initial condition of the particle and naturally admits an information-theoretic interpretation. Because we specify only the mean of the initial distribution $p_i(\bm{x})$, the system may start in an arbitrarily complex state. However, the harmonic trap can only access the portion of this information that is compatible with its fixed shape or position \cite{kolchinsky2021work,gupta2025thermodynamiccoststeadystate}.

To make this more precise, we introduce the M-projection $\pi[p_i]$, defined by \cite{amari2016information}\begin{equation}
    \pi[p_i](\bm{x}) = \underset{\rho\in \mathscr{B}}{\text{argmin}} \: D_\text{KL}(p_i \parallel \rho),
\end{equation} 
where $D_\text{KL}$ is the Kullback--Leibler divergence. This operation projects the initial distribution $p_i(\bm{x})$ onto the space $\mathscr{B}$ of Boltzmann states allowed by the trap manipulation—here, fixed-variance Gaussian densities (often called \emph{shift measures}). By minimizing $D_\text{KL},\pi[p_i]$ is the least-information-loss Gaussian approximation of $p_i$. In our example, $\pi[p_i](\bm{x})$ becomes a Gaussian with center $\bm{q_i}$ and variance determined by the trap shape. One can then show that \begin{equation}\label{eq:DKLinitial}
 k_BT \:   D_\text{KL}(\pi[p_i] \parallel p_\text{eq}) = \frac{1}{2}k(\bm{\lambda}_i - \bm{q}_i)^2
\end{equation}
where $p_\text{eq}$ is the true Boltzmann state of the initial trap. Thus, the first term in Eq.~(\ref{eq:Wqs}) quantifies the \emph{accessible} information within the non-equilibrium initial state that can be transformed into (negative), i.e.\ extracted, work.

\begin{figure}[t!]
    \centering
    \includegraphics[width=\columnwidth]{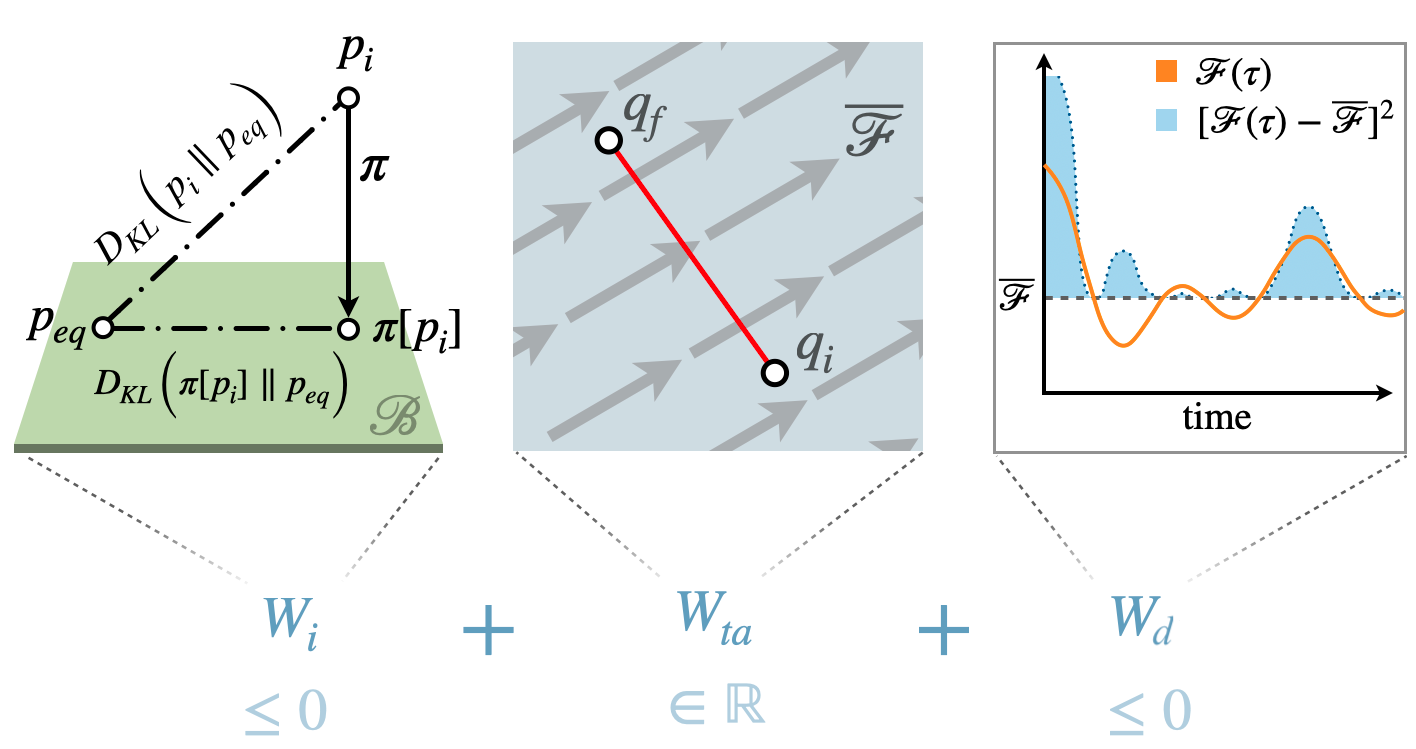}
    \caption{Contributions to the work in the quasistatic limit; a information theoretic work $\mathscr{W}_i$, work due to time-averaged displacement $\mathscr{W}_{ta}$, and work due to force deviations from its time-average $\mathscr{W}_d$. The information theoretic contribution measures the accessible information in the initial non-equilibrium state, which can be extracted as work. The contribution from the time-averaged force can take any sign depending on the direction of forces and desired translation. Finally, deviations around the time-averaged force can be used to extract work, by an amount proportional to the shaded blue area (accumulated square deviations).  }
    \label{fig:terms}
\end{figure}

The two last terms of Eq.~(\ref{eq:Wqs}) are contributions originating in the non-equilibrium driving forces. More precisely, the contribution $\mathscr{W}_{ta}$ originates in the time-averaged force and is simply the work needed to move a particle a distance $\bm{q}_f-\bm{q}_i$ in the presence of $\overline{\bm{\mathscr{F}}}$. The last term is determined by the deviations of the force around the time-averaged mean, and can be written
\begin{align}
    \mathscr{W}_{d} =  - \frac{1}{4 \gamma }  \lim_{t_\text{f}\to \infty}  \int_0^{t_\text{f}} d\tau \bm{\delta\mathscr{F}}(\tau)^2
\end{align}
where $ \bm{\delta \mathscr{F}}(\tau) = \bm{\mathscr{F}}(\tau) - \overline{\bm{\mathscr{F}}}(t_\text{f})$ are the deviations of the force from its time average value. This term encodes how much work can be extracted by utilizing the information about the temporal details of the force. We emphasize here that one can extract more work from time-varying forces than from stationary ones, especially in situations where the deviations from the time-averaged mean are strongly persistent in time.

To gain further insight into this contribution, $\mathscr{W}_d$,  consider a key outcome of the optimization procedure: the Euler--Lagrange equation implies an effective overdamped motion for the particle’s mean position. In the quasistatic limit, this takes the simple form $\gamma \dot {\bm{q}}(t) = {\bm{\delta\mathscr{F}}}(t)/2$. As discussed in the Appendix, this effective equation describes a \emph{free} particle driven by the forces ${\bm{\delta\mathscr{F}}}(t)/2$. One may interpret this as capturing the fast velocity modes of the particle; meanwhile, the slow mode—which transports the particle a finite distance over an infinite time—vanishes in the quasistatic limit, and its associated work is accounted for by $\mathscr{W}_\text{ta}$.

Because the particle effectively experiences the force ${\bm{\delta\mathscr{F}}}(t)/2$, there is a corresponding work contribution with differential increment  $dW_{{\bm{\delta\mathscr{F}}}} = \frac{{\bm{\delta\mathscr{F}}}}{2} d\bm{q}(t)$. From the effective equation of motion, $d\bm{q}(t) = \frac{{\bm{\delta\mathscr{F}}}}{2\gamma} dt$, so the instantaneous power becomes $\dot {W}_{{\bm{\delta\mathscr{F}}}} = \frac{{\bm{\delta\mathscr{F}}}^2}{4\gamma} $. Integrating over the duration of the protocol yields the work done by these forces, which is precisely the extractable work $\mathscr{W}_{d} = - \int_0^{t_\text{f}} d t \dot {W}_{{\bm{\delta\mathscr{F}}}}(t)$. All contributions to the work, Eq.~(\ref{eq:Wqs}), are summarized in Fig.~(\ref{fig:terms}).

{\color{black}
Our above framework leverages information about non-equilibrium forces to enhance energetic efficiency. This is not unlike information engines, which traditionally utilize information through measurement and feedback to rectify fluctuations and extract work \cite{parrondo2015thermodynamics,lutz2015information,goerlich2025experimental}. Our protocols, in some sense, function as automatic information engines. Rather than rectifying fluctuations from a bath, Euler-Lagrange minimization protocols emerge that spontaneously maximize the amount of energy extracted by anticipating and responding to prescribed non-equilibrium dynamics. Below, we explore two case studies that illustrate the versatility of our approach with respect to different force types: externally applied periodic forces and internally generated active self-propulsion forces.
}

\subsection*{Case I: Periodic forces - a minimal automatic information engine}
\begin{figure}[t]
    \centering
    \includegraphics[width = \columnwidth]{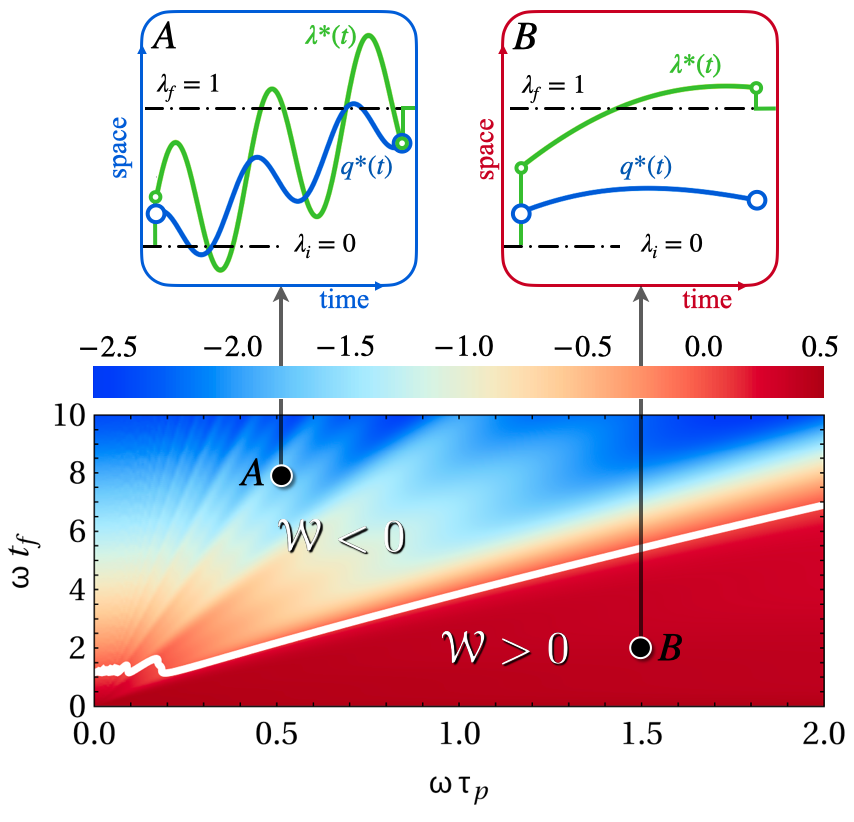}
    \caption{Work as a function of the timescale of force oscillations $\tau_\text{p}$ and protocol duration $t_\text{f}$, showing regions where work must be paid or can be extracted. Insets (points A and B) show optimal protocol (green line) and the associated mean particle trajectory (blue line). Parameters used are $\omega = \gamma = \lambda_f = 1, f_0= -1,  \lambda_i = 0, q_i = \lambda_f/4$.}
    \label{fig:oscill}
\end{figure}

Many energy harvesting solutions are based on periodic forces or motion, such as wave-energy converters, wearable technology where fabrics extract energy from movement, and piezoelectric generators that can charge pacemakers through heartbeats \cite{falcao2010wave,gammaitoni2011vibration,anwar2021piezoelectric,beeby2006energy,azimi2021self}. Here we consider a simple microscopic analogy, which we analyze through the above framework. We consider a Brownian particle effectively confined to one-dimensional movement, exposed to periodic driving forces
\begin{equation}
    \mathscr{F}(t) = f_0\: \sin (t/\tau_\text{p}).
\end{equation}
Here $f_0$ is the amplitude of the force while $\tau_\text{p}$ determines its periodicity. Fig. (\ref{fig:oscill}) shows the work as a function of protocol duration and force periodicity. The associated optimal protocol is shown in regions of both positive and negative work. As is common for optimal protocols, discontinuous jumps are seen in the beginning and final parts of the protocol.

\begin{figure}[t]
    \centering
    \includegraphics[width = \columnwidth]{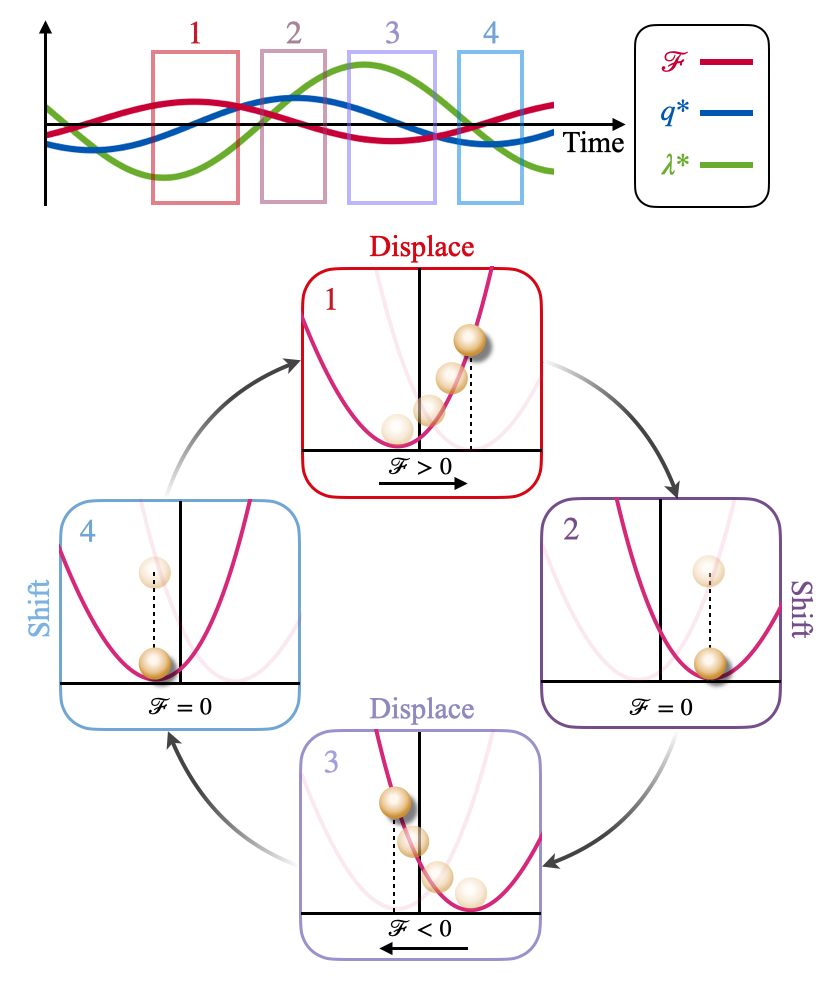}
    \caption{Sketch of the mechanism behind work extraction for periodic forcing. Top panel shows forces (red line), particle position (blue line) and protocol (green line) as a function of time. When the forces are approximately maximal, the particle climbs the potential (panel 1). Once the maximal position is reached, the protocol switches from negative to positive side and catches up to the particle (panel 2). In panels 3 \& 4 this process repeats in the opposite direction.}
    \label{fig:oscill2}
\end{figure}

We see that for sufficiently slow protocols, compared to the forcing period, the optimal control is able to utilize the oscillations such that work can be extracted.  Recall that in Brownian information engines, feedback is used to rectify thermal noise and convert information into work.  Here, the optimal protocol harnesses the dynamic information available and automatically extracts as much work as possible. This is a consequence of precise knowledge of the force at all times.  The protocol consists of repeatedly letting the force move the particle into a high-energy state before shifting the potential accordingly to extract the stored energy as work. Figure (\ref{fig:oscill2}) summarizes the main mechanism behind the work extraction. 

{\color{black}
While the above repeating cycles are specific to the current example, the way in which work is extracted offers insights into more generic situations. Indeed, combining the Euler-Lagrange equation with the equation of motion we have $\dot{ \bm{\lambda}}= \dot {\bm{q}} - \dot{\bm{\mathscr{F}}}/(2k)$. When forces increase in a given direction $\dot{\bm{\mathscr{F}}}>0$ the optimal protocol lets the particle move faster than the trap, $\dot{ \bm{\lambda}} < \dot {\bm{q}}$, lifting the particle to a higher energy state. Once forces start to decrease $\dot{\bm{\mathscr{F}}}<0$ the protocol catches up to the particle $\dot{ \bm{\lambda}} > \dot {\bm{q}}$.
}

The above cycle can, in principle, extract arbitrarily large amounts of work over an indefinitely long protocol. In our example, this follows from Eq.~(\ref{eq:Wqs}), where 
 where $\text{Var}(\mathscr{F}) = f_0^2/2$ remains constant, causing $\mathscr{W}_d \sim t_\text{f} \text{Var}(\mathscr{F})$ dto grow unbounded (in the negative direction) as $t_\text{f} \to \infty$. For finite yet large protocol durations, $t_\text{f} \gg \tau_\text{p}$, the term $\mathscr{W}_d$ dominates, and the work behaves as 
\begin{equation}
    \mathscr{W} = -\frac{f_0^2}{8 \gamma} t_\text{f}
\end{equation}
which is independent of the forcing period $\tau_p$.  

This large potential for work extraction relies on precise initial information about the periodic forces. For example, we can easily extend the model by considering a force $f(t) = f_0 \sin(t/\tau_p +\varepsilon)$ where $\varepsilon$ is a random variable representing our ignorance about phase information. Taking these errors to be normal distributed with zero mean and variance $\sigma_\varepsilon^2$ we obtain the mean force $\mathscr{F}(t) = \exp(-\sigma_\varepsilon^2/2) f_0 \sin(t/\tau_p)$. Hence, incorporating these initial errors effectively rescales the force amplitude, and the work extracted at extensive protocol durations instead takes the form $\mathscr{W} = -\frac{f_0^2\exp(-\sigma_\varepsilon^2) }{8 \gamma} t_\text{f}$. Hence, initial measurement errors can be detrimental to the design of engines,  leading to exponentially reduced work extraction. Notably, while phase errors cause an exponential suppression of work, their degree of suppression remains unchanged over time. Once the phase is incorrectly estimated, the mismatch persists, effectively lowering the force amplitude throughout the entire process.

In the presence of a periodicity error, modeled as, $f(t) = f_0 \sin(t/(\tau+\varepsilon))$, the effective force can be approximated for small $\varepsilon$. When $\varepsilon$ is normally distributed with a small variance $\sigma_\varepsilon$ we use:
\begin{equation}
    \mathscr{F}(t) \approx f_0 \left\langle \sin\left( \frac{t}{\tau_p} -\frac{t}{\tau_p^2} \varepsilon\right) \right\rangle =f_0  \exp\left(-\frac{t^2 \sigma_\varepsilon^2}{2\tau_p^4}\right)\sin\left( \frac{t}{\tau_p} \right). 
\end{equation}
Here, the periodic force experiences a pronounced exponential suppression. In contrast to the case where the error is in the phase, the error in periodicity gives rise to a larger phase mismatch over time, leading to strong destructive interference, which suppresses the effective forces. The work in the quasistatic case can be calculated as before, 
\begin{equation}
    \mathscr{W}_{qs} \approx -\frac{\sqrt{\pi}\tau_p^2f_0^2}{16 \gamma \sigma_\varepsilon}\left(1-e^{-\tau_p^2/\sigma_\varepsilon^2}\right).
\end{equation}
In sharp contrast to the case without errors, work extraction is now bounded. We emphasize that although this approximation may not be quantitatively exact in the quasistatic approximation, it clearly shows how small errors qualitatively affect the extracted work.

\subsection*{Case II: Work extraction from active forces}

{ \color{black}
Active matter represents a compelling category of non-equilibrium systems, where mechanisms for work extraction have been investigated recently \cite{di2010bacterial,reichhardt2017ratchet,gupta2023efficient,paneru2022colossal,malgaretti2022szilard,cocconi2023optimal,cocconi2024efficiency,garcia2024optimal,schüttler2025activeparticlesmovingtraps,baldovin2023control,davis2024active}. In contrast to the previous example of an external oscillating force, the forces that drive active particles are stochastic and internally generated. Here we apply our general framework to several active matter scenarios, and compare the resulting work extraction. }

\begin{figure}[t]
    \centering
    \includegraphics[width=8.66cm]{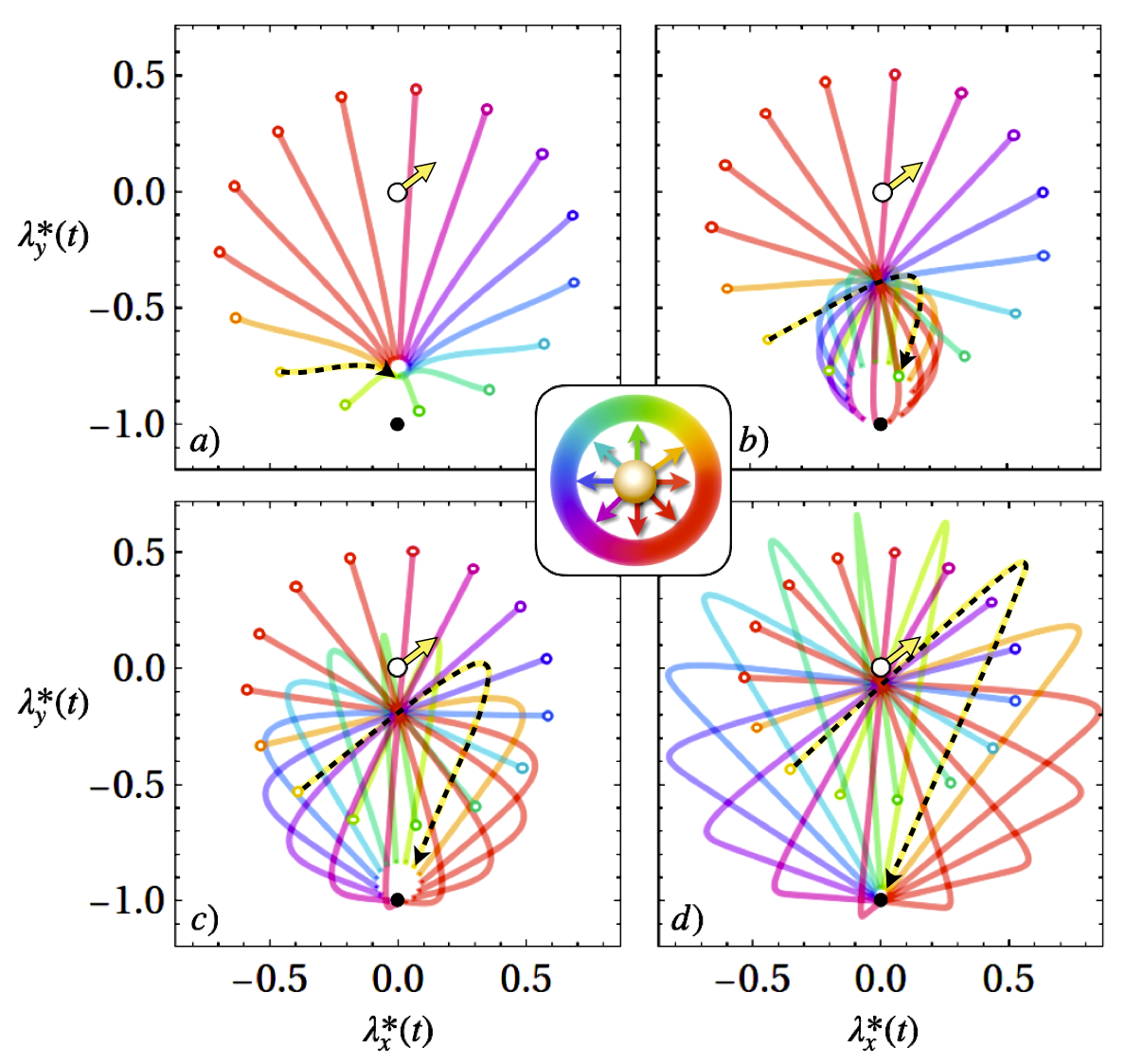}
    \caption{ Optimal protocol for an ABP starting at the origin with initial orientations $\phi_0$ indicated by the middle inset. In units of the persistence length $\ell_p = f_0/(\gamma D_r)$ the protocol starts at the origin together with the particle (small black circle) and ends at $(0,-1)$ (small black dot). Panels a) - d) respectively show $D_r t_\text{f} = 2,D_r t_\text{f} = 5,D_r t_\text{f} = 10$ and $D_r t_\text{f} = 30$. In all cases, the potential starts and ends with a discontinuous jump. Dashed line highlights a protocol associated with the initial orientation given by the yellow arrow.  Parameters used are $D_r \tau_R = 2, f_0 = \omega = \gamma =1$.}
    \label{fig:ABPprot}
\end{figure}

We consider an active particle in two dimensions, modeled as an active Brownian particle, obeying the stochastic equation of motion 
\begin{align}
     \dot {\bm{x}} (t) &= \frac{f_0}{\gamma} \hat{\bm{n}}(t)  - \omega [\bm{x}(t) - \bm{\lambda}(t) ] + \sqrt{2 D} \bm{\xi}(t)
\end{align}
where $f_0$ is the constant magnitude of the active self-propulsion force, and $\hat{\bm{n}}(t)  = [\cos \phi(t),\sin\phi(t)]$ determines the direction of propulsion. Persistence is expected to play a key role in the energy extraction in any active system, motivating us to consider not only reorientations driven by white noise, as is usual for ABPs, but a slightly more realistic situation where finite-time correlations are present. In the simplest case, one can consider orientations driven by an Ornstein-Uhlenbeck noise 
\begin{align}
    \dot \phi(t) &= \Omega(t)\\
   \dot  \Omega(t) &= -\frac{1}{\tau_R} \Omega + \frac{\sqrt{2D_r}}{\tau_R}  \eta(t),
\end{align}
where $D_r$ is a rotational diffusion coefficient and $\tau_R$ a relaxation timescale associated with the rotational dynamics. Such models have been considered in the past to include memory effects in the orientational dynamics, resulting for example from misalignments with the instantaneous propulsion direction or inertial effects \cite{weber2011active,ghosh2015communication,scholz2018inertial,lowen2020inertial}. When $\Omega$ is stationary, the distribution of $\phi(t)$ is known to be Gaussian with mean $\phi_0$ and variance $\sigma_\phi^2(t) = 2 D_r (t-\tau_R (1-e^{-t/\tau_R}))$. This results in a mean force
\begin{equation}
    \bm{\mathscr{F}}(t|\phi_0) = \langle f_0 \hat {\bm{n}}(t) |\phi_0\rangle = f_0 \begin{bmatrix}
\cos(\phi_0) e^{-\frac{1}{2}\sigma_\phi^2(t)}\\
\sin(\phi_0) e^{-\frac{1}{2}\sigma_\phi^2(t)}
\end{bmatrix},
\end{equation}
conditioned on knowing the initial force, i.e., propulsion direction, as can be obtained from a measurement \cite{garcia2024optimal}. From this, the optimal protocol and the associated work can be calculated. Calculations of both protocol and associated work requires calculating the time-integrated mean and variance of the above force. This can be done exactly, detailed in the appendix. The protocol $\bm{\lambda}^*(t)$ is then calculated directly from Eq.(\ref{eq:lambdaeq}-\ref{eq:lambdaneq}), with the results shown in Fig.(\ref{fig:ABPprot}). We see that short protocol durations (e.g., panel a) results in protocols with large initial jump and almost linear dragging. Longer protocol durations however (e.g., panel d) shows smaller jumps and a curved protocol trajectory that utilizes the particle persistence to lower the energetic cost, or even allows work to be extracted. Generally, the protocol makes a jump to a position behind the particle,  and follows the particle's initial direction of motion for a while. This slows the particle down and after the particle's persistence is lost due to rotational noise, the protocol drags the particle back to the target location.

\begin{figure}[t]
    \centering
    \includegraphics[width=8.65cm]{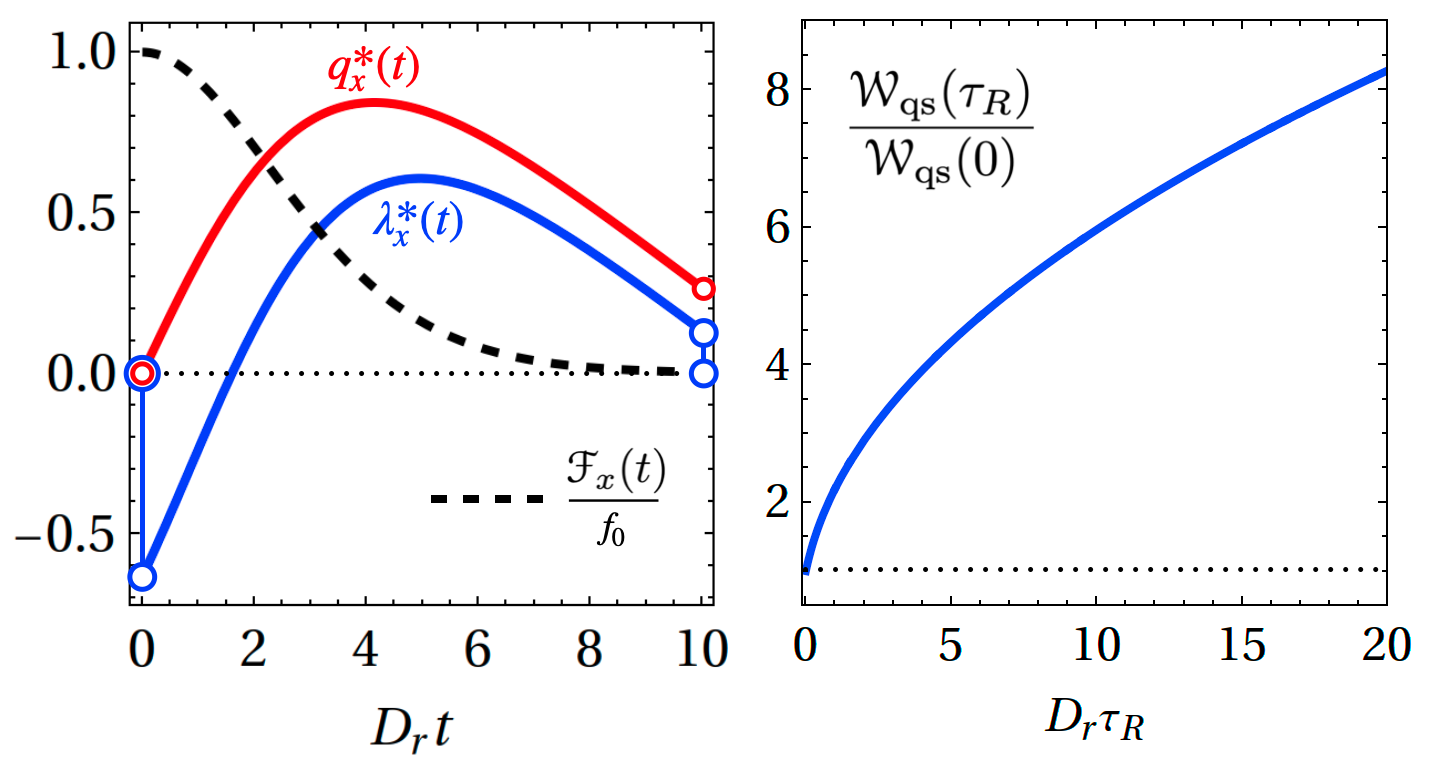}
    \caption{ Protocol for an ABP starting and ending at the origin with $\lambda_x(0) = \lambda_{x,f} = q_x (0) = 0$. $D_r\tau_R = 5,D_r t_\text{f} = 10, f_0 = \omega = \gamma =1$.}
    \label{fig:ABP1}
\end{figure}


\begin{table*}[t]
\begin{tabular}{ |>{\centering\arraybackslash}m{1.7cm}||>{\centering\arraybackslash}m{4.3cm}|>{\centering\arraybackslash}m{4cm}|>{\centering\arraybackslash}m{3.4cm}|>{\centering\arraybackslash}m{3.1cm}|  }
 \hline
 \multicolumn{5}{|c|}{ Active particle dynamics and quasistatic work extraction bound}\\[7pt]  \hline
 $$\text{Model}$$ & Fractional angular noise $$ \phi(t)  \sim \text{fBm}(H) $$ &  Random rotational diffusion  $$D_r \sim  \frac{1}{\overline{D}} e^{-D_r/\overline{D}}$$ & Chiral ABP  $$ \dot \phi(t) = \omega_c +\sqrt{2 D_r} \eta(t)$$& Accelerating ABP $$\bm{f}(t) = f_0 (t/\tau)^\alpha \hat{ \bm{n}}(t)$$ \\
 \hline \\[-14pt]

$$\text{Effective}$$ $$\text{force } \bm{\mathscr{F}}(t)$$   &  $$ f_0 \hat{\bm{n}}_0 e^{- D_H  t^{2H}}$$   & $$\frac{f_0\hat{\bm{n}}_0 }{1+\overline{D}t}  $$ &   $$ f_0 \begin{bmatrix}
     \cos(\phi_0+\omega t)\\
     \sin(\phi_0+\omega t)
 \end{bmatrix} e^{-D_r t} $$ & $$ f_0\hat{\bm{n}}_0 (t/\tau)^\alpha e^{-D_r t} $$\\

 \hline \\[-23pt]
  $$\text{Quasistatic}$$  $$\text{work}$$    &  $$ \frac{\mathscr{W}_\text{qs} (H)}{\mathscr{W}_\text{qs} (1/2)}= \frac{D_{1/2} \Gamma(1 +\frac{1}{2H})}{2^{\frac{1}{2H}-1}D_H^{\frac{1}{2H}}  }$$   & $$\mathscr{W}_\text{qs} (\overline D = D_r)=2 \mathscr{W}_\text{qs}^\text{ABP}$$ &   $$\mathscr{W}_\text{qs}=\mathscr{W}_\text{qs}^\text{ABP}$$ & $$\frac{ \mathscr{W}_\text{qs}}{ \mathscr{W}_\text{qs}^\text{ABP}} = \frac{\Gamma(1+2\alpha)}{4^\alpha (D_r \tau)^{2\alpha}}$$\\

 \hline
\end{tabular}
\caption{Various active particle models along with the quasistatic work extraction given by Eq. (\ref{eq:Wqs}). Cases considered: 1) Fractional Brownian orientations characterized by a Hurst exponent $H$ such that $H \in (0,1/2)$ gives anti-correlation reorientations, and $H\in (1/2,1)$ correlation reorientations. At $H = 1/2$ normal ABP dynamics is recovered with $D_{1/2} = D_r$. 2) Random rotational diffusion, whereby the rotational diffusivity is random and exponentially distributed with mean $\overline{D}$. 3) Chiral ABPs where a mean angular velocity $\omega_c$ is included. 4) Accelerating ABPs where the self-propulsion force depends on time as a power-law with exponent $\alpha \in (-1/2,\infty)$. A direct comparison with the traditional ABP result $\mathscr{W}_\text{qs}^\text{ABP} = - f_0^2/(8\gamma D_r)$ is given in all cases.} \label{tab:models}
\end{table*}

In terms of work extraction, we can immediately obtain the quasistatic bound $\mathscr{W}_\text{qs}$ from Eq.(\ref{eq:Wqs}). Without loss of generality, we set $\phi_0 =0$, such that $\mathscr{F}_y = \lambda_y^* (t) = 0$ by symmetry, and only the $x$-components of the protocol needs to be considered. This results in 
\begin{equation}
   \frac{\mathscr{W}_\text{qs}(\mathfrak{D}_0)}{\mathscr{W}_\text{qs}(0)} = \frac{e^{2 \mathfrak{D}_0}}{(2\mathfrak{D}_0)^{2 \mathfrak{D}_0-1}} \left[ \Gamma(2\mathfrak{D}_0) - \Gamma(2\mathfrak{D}_0|2\mathfrak{D}_0) \right]
\end{equation}
which is expressed solely in terms of the dimensionless number $\mathfrak{D}_0 = D_r\tau_R$, introduced as a \emph{delay number} in Ref. \cite{scholz2018inertial}. Here we normalized the work in terms of its $\tau_R \to 0$ value 
\begin{equation}
    \mathscr{W}_\text{qs}(0) \equiv \mathscr{W}_\text{qs}^\text{ABP} = - f_0^2/(8D_r \gamma)
\end{equation}
which is the bound for the normal ABP model with Gaussian white noise driving the orientations. 
Figure (\ref{fig:ABP1}), left panel, shows the optimal protocol and the associated mean particle position for a persistent active particle with initial orientation along the positive $x$-axis. We see that the optimal protocol has an initial jump behind the particle. Although this comes at a high work (as reported also in Ref. \cite{garcia2024optimal}) this allows the protocol to extract work from the persistent nature of the particle during the remainder of the protocol. Figure (\ref{fig:ABP1}), right panel, shows the work in the quasistatic limit, monotonically growing as a function of the delay number $\mathfrak{D}_0$, indicating the positive effect of finite-time angular correlations. We emphasize that while the ratio $\mathscr{W}_\text{qs}(\mathfrak{D}_0)/\mathscr{W}_\text{qs}(0)$ is positive, the work $\mathscr{W}_\text{qs}(\mathfrak{D}_0)$ is strictly negative for all values of the delay number, showing that work is effectively extracted.

The quasistatic work bound given by Eq. (\ref{eq:Wqs}) can easily be evaluated for a wide range of other active particle models as well. In table \ref{tab:models} we list four other active particle models and the quasistatic work associated with their optimal protocols. More precisely, we consider active particles driven by fractional Brownian noise \cite{gomez2020active}, a random rotational diffusion model, chiral particles \cite{van2008dynamics,lowen2016chirality} and particles that accelerate or decelerate in time \cite{babel2014swimming}. For the random rotational diffusivity model with mean fixed to the rotational diffusion of a pure ABP we find double the potential for work extraction, while chirality does not affect the bound. For fractional angular noise and accelerating ABPs the work extraction may be higher or lower than the pure ABP case depending on parameters.

\section*{Discussion}

{ \color{black}
As technology trends toward miniaturization, realistic devices will increasingly be exposed to complex, dynamic environments characterized by time-dependent forces and fluctuations. In such settings, optimal control is not merely a theoretical pursuit, but essential for efficient and reliable control and energy harvesting. Here, exact results valid for arbitrary time-dependent driving forces has been derived using methods from finite-time stochastic thermodynamics and optimal control theory. We showed that the quasistatic work associated with optimal protocols naturally decomposes into three contributions; i) an information-geometric term representing how the information contained in an initial non-equilibrium state can be converted to work, ii) the work associated with slowly dragging a particle in the presence of time-averaged forces, and iii) additional work extraction facilitated by protocols responding to fast dynamical modes.  In the presence of forces with a temporal evolution leading to a sufficiently slowly decaying time-integrated variance, unbounded work extraction can be obtained if given enough time. Our work offers a broad perspective on how time-dependent forces can be optimally utilized to extract work.

We illustrated our framework by considering two case scenarios; deterministic periodic forces, and stochastic active self-propulsion forces. In the case of periodic forces, the extracted work grows linearly with the protocol duration, and the optimal protocol behaves like an automatic information engine. Additionally, we explored the realm of active matter by considering in detail an active particle with finite-time angular correlations. Here the extracted work was found to grow monotonically with the correlation timescale, underscoring the significance of properly accounting for these correlations. Finally, a wide range of other active particle models were considered, and their quasistatic work bounds compared.

A future application of the principles investigates here could be in the design of microscopic robots for medical applications such as targeted drug delivery \cite{manjunath2014promising,saadeh2014nanorobotic}. Additionally, experimental techniques in finite-time thermodynamics now allow us to explore questions of optimal control and energy extraction. These methods offer a promising pathway to validate the theoretical findings presented here.

Our work highlights the role of nonequilibrium forces and fluctuations in extracting work, drawing inspiration from biological systems that naturally harvest energy from nonequilibrium environments. The problem of maximizing work extraction is crucial not only for understanding energy extraction  and transduction in living systems but also for advancing micro-technological applications \cite{hill2012free, pinero2023universal, kolchinsky2025maximizing}. By exploring these principles, we highlight fundamental limits to energy harvesting from dynamical forces, relevant in both biological and technological contexts.

}

\acknowledgements
K.S.O acknowledges support from the Alexander von Humboldt foundation. Y.R. acknowledges support from the Israel Science Foundation (grants No. 385/21). Y.R. and R.G. acknowledge support from the European Research Council (ERC) under the European Union’s Horizon 2020 research and innovation program (Grant Agreement No. 101002392). R.G. acknowledges support from the Mark Ratner Institute for Single Molecule Chemistry at Tel Aviv University. H.L acknowledges support by the Deutsche Forschungsgemeinschaft (DFG) within the project LO 418/29-1.

\appendix
\section{Derivation of the optimal protocols}

Here, we derive the optimal protocol for arbitrary driving forces, as well as the associated mean particle trajectories and the thermodynamic work. We define the mean particle position as $\bm{q}(t) = \langle \bm{x}(t) \rangle$. The mean particle position evolves as
\begin{equation}
   \gamma \dot {\bm{q}}(t)= - k [\bm{q}(t)-{\bm{\lambda}}(t)] + {\bm{\mathscr{F}}}(t) 
\end{equation}
We let $\omega = k/\gamma$, so that the equation of motion takes the form 
\begin{equation}
    \dot {\bm{q}}(t)= - \omega [\bm{q}(t)-{\bm{\lambda}}(t)] + {\bm{\mathscr{F}}}(t)/\gamma 
\end{equation}

For a harmonic control trap we have the mean work
\begin{align}
    \mathscr{W} &= k\int_0^{t_\text{f}} dt \dot {\bm{\lambda}}(t) \cdot \left[ \bm{\lambda}(t)-\bm{q}(t)\right]\\
    &=  \int_0^{t_\text{f}} dt \: \dot {\bm{\lambda}}(t) \cdot \left[\gamma  \dot {\bm{q}}(t) -{\bm{\mathscr{F}}}(t) \right]
\end{align}
where we used the equation of motion. Taking a further derivative of the mean equation of motion, we can also write
\begin{equation}
 \ddot {\bm{q}}(t)= - \omega \dot{\bm{q}}(t) + \omega \dot{\bm{\lambda}}(t) + \dot{{\bm{\mathscr{F}}}}(t) /\gamma
\end{equation}
which we can use to eliminate $\omega \dot {\bm{\lambda}}(t)$ in the expression for the work. This gives
\begin{equation}
    \mathscr{W} =  \frac{\gamma}{\omega}\int_0^{t_\text{f}} dt \left[ \ddot {\bm{q}}(t) + \omega\dot {\bm{q}}(t) - \dot {\bm{\mathscr{F}}}(t)/\gamma   \right]  \cdot \left[ \dot {\bm{q}}(t) - {\bm{\mathscr{F}}}(t)/\gamma \right]
\end{equation}
Many of these terms can be written as total time derivatives. For example, $\dot {\bm{q}} \cdot  \ddot{\bm{q}} = \frac{1}{2} d (\dot {\bm{q}}^2)/dt $. Proceeding similarly, we have

\begin{align}
   \mathscr{W} &= \frac{1}{2k}[(\gamma \dot {\bm{q}})^2(t)]_0^{t_\text{f}}  +  \frac{1}{2k}[{\bm{\mathscr{F}}}^2(t)]_0^{t_\text{f}}  -\frac{1}{\omega}[\dot {\bm{q}}(t) \cdot {\bm{\mathscr{F}}}(t)]_0^{t_\text{f}} \\
    &+  \int_0^{t_\text{f}} dt \mathscr{L}(t, \bm{q},\dot {\bm{q}})
\end{align}

where we introduced the Lagrangian $\mathscr{L}(t, \bm{q},\dot {\bm{q}}) = \gamma\dot {\bm{q}}(t) \cdot [ \dot {\bm{q}}(t) -{\bm{\mathscr{F}}}(t)/\gamma ]$. The corresponding Euler-Lagrange equation reads
\begin{align}
    \ddot {\bm{q}}(t) = \frac{\dot {\bm{\mathscr{F}}}(t)}{2\gamma}
\end{align}
which can easily be solved by
\begin{align}\label{eq:q_inter}
     \bm{q}(t) = \bm{q}_i + \bm{\varphi} t + \frac{1}{2\gamma}\int_0^{t} dt' \bm{\mathscr{F}}(t')
\end{align}
where we used the initial condition $ \bm{q}(0) = \bm{q}_i$.  To proceed, we must identify the unknown constant $\bm{\varphi}$, which is typically done by minimizing the work with respect to $\bm{\varphi}$ \cite{schmiedl2007optimal}. For this, we must first write the explicit expression for the work, which we do by using the equations of motion to determine the boundary conditions for the particle velocity:
\begin{align}
    \dot {\bm{q}}(0) &= \omega(\bm{\lambda}_i-\bm{q}_i) + \bm{\mathscr{F}}(0)/\gamma \\
    \dot {\bm{q}}(t_\text{f}) &= \omega(\bm{\lambda}_f-\bm{q}(t_\text{f})) + \bm{\mathscr{F}}(t_\text{f}) /\gamma
\end{align}
The work then explicitly reads
\begin{align}\label{eq:w_final}
    \mathscr{W} &= \frac{1}{2k}[ (k(\bm{\lambda}_f -  \bm{q}(t_\text{f}) )+\bm{\mathscr{F}}(t_\text{f}))^2] \nonumber\\
    &-\frac{1}{2k} [(k(\bm{\lambda}_i -  \bm{q}_i )+\bm{\mathscr{F}}(0))^2]  +  \frac{1}{2k}[\bm{\mathscr{F}}(t_\text{f})^2 -\bm{\mathscr{F}}(t_0)^2] \nonumber \\
    & - \frac{1}{k}[ (k(\bm{\lambda}_f -  \bm{q}(t_\text{f}) )+\bm{\mathscr{F}}(t_\text{f})) \cdot  \bm{\mathscr{F}}(t_\text{f})] \nonumber \\
    &+\frac{1}{k} [ (k(\bm{\lambda}_i -  \bm{q}_i )+\bm{\mathscr{F}}(t_0))\cdot\bm{\mathscr{F}}(t_0)] \\
    & + \gamma \bm{\varphi}^2 t_\text{f}  - \gamma\int_0^{t_\text{f}} dt   \left(\frac{\bm{\bm{\mathscr{F}}}(t)}{2\gamma}\right)^2 
\end{align}
We note that there is here also a dependence on $\bm{\varphi}$ through terms containing $\bm{q}(t_\text{f})$, from Eq. (\ref{eq:q_inter}). From this expression, the unknown parameter $\bm{\varphi}$ can be determined by minimization $\partial_{{\bm{\varphi}}_*}\mathscr{W} =0$, resulting in

\begin{align}\label{eq:phi}
    \bm{\varphi}_* = \omega\frac{\bm{\lambda}_f-\bm{q}_i}{2 + \omega t_\text{f} } - \frac{ \omega}{ 2 + \omega t_\text{f}} \int_0^{t_\text{f}} dt \frac{\bm{\mathscr{F}}(t)}{2\gamma}
\end{align}
This gives the mean particle trajectory, which in turn through the equation of motion can be used to find the optimal protocol. It takes the form
\begin{align}
    \bm{\lambda}_*(t) &=  \bm{\lambda}_\text{eq}(t)   + \bm{\lambda}_\text{neq}(t)
\end{align}
where
\begin{equation}
    \bm{\lambda}_\text{eq}(t) = \bm{q}_i + \frac{1+\omega t}{2+\omega t_\text{f}}(\bm{\lambda}_f-\bm{q}_i)
\end{equation}
and 
\begin{align}
\bm{\lambda}_\text{neq}(t) = \int_0^{t} dt' \frac{\bm{\mathscr{F}}(t')}{2\gamma}  -   \frac{1+ \omega t}{2+\omega t_\text{f}}   \int_0^{t_\text{f}} dt \frac{\bm{\mathscr{F}}(t)}{2\gamma} -  \frac{\bm{\bm{\mathscr{F}}}(t)}{2k} 
\end{align}

Combining results, the work can be written 

\begin{align}\label{eq:w_final}
    \mathscr{W}  &=     \frac{k}{2}[ (\bm{\lambda}_f -  \bm{q}(t_\text{f}) )^2] 
    -\frac{k}{2} [(\bm{\lambda}_i -  \bm{q}_i )^2]   \\
    & + \left(  k \frac{\bm{\lambda}_f-\bm{q}_i}{2 + \omega t_\text{f} } - \frac{k}{ 2 + \omega t_\text{f}} \int_0^{t_\text{f}} dt \frac{\bm{\mathscr{F}}(t)}{2\gamma} \right)^2  \frac{ t_\text{f} }{\gamma} \nonumber \\
    &- \gamma \int_0^{t_\text{f}} dt   \left(\frac{\bm{\bm{\mathscr{F}}}(t)}{2\gamma}\right)^2 
\end{align}

\section{Effective particle dynamics in the quasistatic limit}
The particle dynamics is a result from many competing effects. Indeed, in addition to white noise, the particle experiences arbitrary time-dependent forces as well as a harmonic potential with mobile center. In the optimal protocol however, the trap center also accounts for the forces. This simplifies the effective dynamics of the particle significantly. 

The mean particle position within the optimal protocol can be studied from the Euler-Lagrange equation $ \gamma \ddot {\bm{q}}(t) = \dot{\bm{\mathscr{F}}}(t)/2$, which we derived above. Integrating once, we find $ \gamma \dot {\bm{q}}(t) = {\bm{\mathscr{F}}}(t)/2 + \gamma\bm{\varphi}(t)$ where $\bm{\varphi}(t)$ is an additional effective force experienced by the particle on top of ${\bm{\mathscr{F}}}(t)/2$. This is given by Eq. (\ref{eq:phi}). In the quasistatic limit, $\gamma \bm{\varphi}(t) =-\overline{\bm{\mathscr{F}}}(t)/2$. The resulting equation for the mean position is simply $\gamma \dot{\bm{q}} = \frac{\bm{\delta \mathscr{F}}}{2}$. Remarkably, no direct effect of the potential can be seen, and effectively the particle is freely moving under the effect of the force $\frac{\bm{\delta \mathscr{F}}}{2}$. Through optimization, the effect of the time-averaged component of the force cancels the effect of the moving harmonic trap. This equation may at first seem paradoxical, since for example in the case of a constant force it states that there is no net movement. Yet, the protocol is designed to transport the particle between two prescribed locations. However, since we work in the quasistatic limit, a finite transport distance performed in infinite time will not contribute to the velocity $\dot{\bm{q}}$. Hence the equation $\gamma \dot{\bm{q}} = \frac{\bm{\delta \mathscr{F}}}{2}$ is naturally interpreted as describing the fast modes of the particle on top of the slow mode that coupled time-averaged force with transport distance.

\section{Time-integrated moments of the active forces}
In the main text we considered an active particle which experiences a mean self-propulsion force
\begin{equation}
    \bm{\mathscr{F}}(t|\phi_0) = \langle f_0 \hat {\bm{n}}(t) |\phi_0\rangle = f_0 \begin{bmatrix}
\cos(\phi_0) e^{-\frac{1}{2}\sigma_\phi^2(t)}\\
\sin(\phi_0) e^{-\frac{1}{2}\sigma_\phi^2(t)}
\end{bmatrix}
\end{equation}
In order to calculate optimal protocols and the associated work, we need the first two time-integrated moments
\begin{equation}
    \overline{\mathscr{F}_\alpha^n} = \frac{1}{t}\int_0^t d t´\mathscr{F}_\alpha^n(t´)\:,\: n=1,2,
\end{equation}
where $\alpha \in \{x,y\}$ determines the spatial directions. Since the time-dependence only comes from the exponential factor $e^{-\frac{1}{2}\sigma_\phi^2(t)}$ with $\sigma_\phi^2(t) = 2 D_r (t-\tau_R (1-e^{-t/\tau_R}))$ we only consider this integral in this appendix. We have to perform the integral
\begin{equation}
    I_n = \int_0^t dt´ \left(e^{- D_r (t-\tau_R (1-e^{-t/\tau_R}))}\right)^n
\end{equation}
We start with $I_1$, for which we perform the substitution $u \equiv e^{-t/\tau_R}$, transforming the integral into
\begin{equation}
    I_1 = \tau_R \int_{e^{-t/\tau_R}}^1 du \: u^{D_r\tau_R-1} e^{ D_r\tau_R (1-u)}
\end{equation}
This can be performed exactly using the incomplete Gamma function, resulting in 
\begin{align}
    I_1 = \frac{\tau_R e^{-D_r \tau_R}}{(D_r \tau_R)^{D_r \tau_R}} \bigg [ &\Gamma\left(D_r\tau_R \bigg | D_r\tau_R e^{-t/\tau_R}\right) \nonumber \\ 
    -&\Gamma\left(D_r\tau_R \bigg | D_r\tau_R \right) \bigg]
\end{align}
Since the integrand is purely exponential, we also immediately see that for the second moment, it sufficed to observe that $I_2(D_r) = I_1(2 D_r)$.

\begin{figure}
    \centering
    \includegraphics[width=8.5cm]{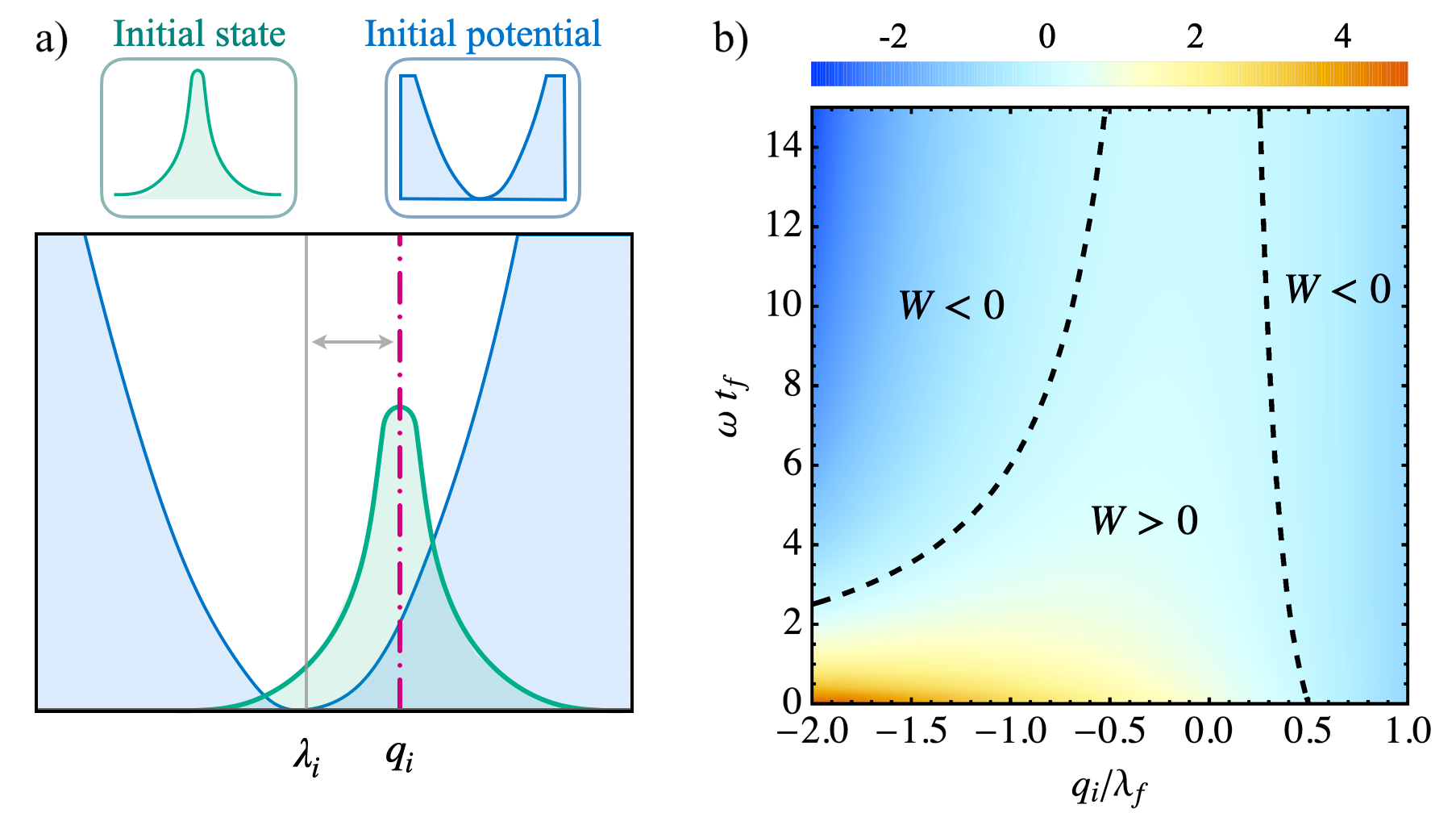}
    \caption{a) Sketch of the initial state, which is assumed to be a non-Boltzmann state with mean $q_i \neq \lambda_i$. b) Work, measured in units of $\frac{k}{2} \lambda_f^2$, as a function of protocol duration and initial mean position.}
    \label{fig:nodrive}
\end{figure}

\section{Baseline example: no forces}

As a baseline example, we consider the case of no forces. At the initial time $t=0$ we assume that the particle has been initialized away from the Boltzmann state associated with the trap, so that $q_i \neq \lambda_i$. See Fig.(\ref{fig:nodrive}a) for a sketch. Without loss of generality, we set $\lambda_i =0$. Since there are no external driving forces in this case, the optimal protocol is simply $\lambda^* = \lambda_\text{eq}$. The effect of this initialization is seen both in the optimal protocol and in the associated work. The protocol has discontinuous jumps both at its beginning and end, a common feature of such protocols \cite{schmiedl2007optimal}. At the beginning the jump is $\Delta \lambda_i = \lambda^*(0)-\lambda_i$,  and at the end $\Delta \lambda_f = \lambda_f-\lambda^*(t_\text{f})$. From the main results, we have 
\begin{equation}
     \Delta \lambda_i = \Delta \lambda_f + q_i-\lambda_i
\end{equation}
with $\Delta \lambda_f = (\lambda_f-q_i)/(2+\omega t_\text{f})$. Hence, jumps are asymmetric ($\Delta \lambda_i  \neq \Delta \lambda_f $) if the particle is initialized away from the trap center. Furthermore, while the discontinuous jumps typically vanish in the quasistatic limit, here the non-equilibrium initial state leads to an initial jump of size $q_i-\lambda_i$ even in the quasistatic limit. This initial jump mimics instantaneous equilibration protocols where the protocol instantaneously changes to conform to the initial state \cite{esposito2011second,parrondo2015thermodynamics}.

The initial non-equilibrium state also has implications for finite-time protocols. The thermodynamic work takes the form
\begin{equation}
    \mathscr{W} =  k\frac{(\lambda_f-q_i)^2}{2+\omega t_\text{f}} - \frac{1}{2}k (\lambda_i-q_i)^2.
\end{equation}

The phase space over which positive and negative regions can be identified is shown in Fig.(\ref{fig:nodrive}b). The zero-work lines, at fixed $t_\text{f}$, can be predicted exactly and are given by the two initial positions 
\begin{equation}
    \frac{q_i^\pm}{\lambda_f} = \frac{1}{1\pm \sqrt{1+\frac{\omega t_\text{f}}{2}}}
\end{equation}
While in the quasi-static limit, any initial deviation from the trap center results in work extraction, for finite-time protocols a sufficiently large displacement is needed.

\section*{References}


%

\end{document}